\documentclass{article}
\usepackage{amsmath,amssymb,bm,cite}
\def\rd{\mathrm{d}}
\def\rD{\mathrm{D}}

\def\rf{\mathrm{f}}
\def\rr{\mathrm{r}}
\def\rs{\mathrm{s}}
\def\ra{\mathrm{a}}
\def\re{\mathrm{e}}
\def\rm{\mathrm{m}}

\begin{document}

\noindent {\large \textbf{Relativistic frequency shifts in a rotating waveguide}}

\vspace{2\baselineskip}\noindent
Mohammad~Khorrami~{\footnote{mamwad@alzahra.ac.ir}},
Amir~H.~Fatollahi~{\footnote{fath@alzahra.ac.ir}},
Ahmad~Shariati~{\footnote{a.shariati@alzahra.ac.ir}}

\vspace{2\baselineskip}\noindent
{\it Department of Physics, Faculty of Physics and Chemistry,
Alzahra University, Tehran 1993891167, Iran}

\vspace{1\baselineskip}
\begin{abstract}
\noindent
A photon source is located in a rotating waveguide.
An absorber with a sharp absorbing frequency absorbs some of the emitted
photons. This decreases the number of photons which are detected by a detector.
The frequency (energy) spectrum measured in the absorber depends
on the setup (the positions of the source and the detector),
and this spectrum determines the amount of absorption.
The shifts are calculated for different configurations.
Among other things, it is seen that even if the source and
the absorber are at rest (while the waveguide is rotating),
there is a nonzero frequency shift.
\end{abstract}

\vspace{2\baselineskip}


\section{Introduction}
Relativity is one of the cornerstones of today's theoretical physics.
During more than a century, several experiments have been designated to
test the theory or determine its domain of validity. Among these are
tests based on the frequency shift. First order shifts are easier
to measure, but there also cases where there is no first order shift
and the leading effect is second order. There are also gravitational
shifts, resulting from the difference in the gravitational potential.
Among these is the Pound-Rebka experiment \cite{poundrebka}
of the gravitational frequency shift based on the M\"ossbauer spectroscopy
\cite{moss}. The M\"ossbauer spectroscopy can also be used for
shifts which are caused by motion. Important examples of this are
the M\"ossbauer rotor experiments \cite{hay,cham,kund}. In these
experiments, the effect of frequency shift on the absorption rate of
a rotating absorber is measured. In the case that the source is at the
center of the rotating disc and the absorber is rotating,
the frequency which is measured by the absorber is blue shifted
by the Lorentz factor corresponding to the absorber \cite{kund,pauli}.
The M\"ossbauer rotor experiments have been extensively studied. 
Some more recent studies are found in \cite{frg,ben1,cor}.
There have also been attempts to interpret such phenomena  
in terms of general-relativistic effects. Some examples are
\cite{mal,rug,ktom,ben2,ben3}.

Motion can affect other properties of photons as well. A review of some
classical and quantum aspects of such effects has been presented in \cite{tiw}.
Among these are the frequency (Doppler) shifts caused by linear or rotational
motions, as well as the so called angular momentum Doppler effect. This effect,
introduced in \cite{gar}, involves the change of the angular momentum of
a photon as a result of its interaction with a moving medium.
This has been further discussed in \cite{fpw,gib}.

Inspired by the M\"{o}ssbauer rotor experiments performed so far,
here the frequency shift is calculated, which resulted from the rotation
of the waveguide in which the emitted photons move. The calculation
is performed for several possible configurations. In the setup, there is
a waveguide which is on a diagonal of a rotating disc, and rotates
with it. A photon source is in the waveguide and rotates with the disc.
An absorber absorbs some of the photons, and after that there is a detector.
The presence of the absorber deceases the number of photons which reach the detector.
The absorption probability (fraction) depends on the frequency of the photons
as measured by the absorber. The frequency shifts are derived using both ray
and particle methods, leading to equivalent results (as expected). In
the configurations which are studied here, the source rotates with the waveguide;
while the absorber either rotates with the waveguide, or is stationary.
A special case of these configurations is when the source is at the center of
the rotating disc, hence stationary. It is seen that in that case, when
the absorber rotates with the disc it measures a frequency which is
blue shifted by the Lorentz factor corresponding to the absorber \cite{kund}.
But when the absorber is stationary, it still measures a blue shifted frequency,
this time corresponding to the square of the Lorentz factor corresponding to
the edge of the disc. This is in spite of the fact that in this case both
the source and the absorber are at rest.

The scheme of the paper is the following. In section 2, the setup and introduced
and the frequency shift (the energy change) is calculated for several
possible setups. In section 3, the number of detected particles is calculated
for those setups. Section 4 is devoted to the concluding remarks.
The relations governing the collision of photons with a wall are
presented in the appendix.
\section{The frequency shift (the energy change)}
A source is located at a disc of radius $r$, in a radial waveguide.
The disc and the radial waveguide rotate (according to the rest frame)
with the angular velocity $\omega$. The distance of the source from
the disc center is $\rho_\rs$. An absorber; either is
in the same waveguide, rotating with the disc and at a distance $\rho_\ra$
from the disc center; or is outside the disc and stationary
(according to the rest frame). A detector detects
the particles (photons) which have not been absorbed by the absorber.
$(\rho,\phi)$ denote the polar coordinates, with the origin at the disc center.
Unprimed (primed) quantities are those which are measured according
to the rest frame (the absorber). The red shift from the source to the absorber
is denoted by $z$:
\begin{align}
z&=\frac{\nu'_\rs}{\tilde\nu}-1.
\intertext{Where $\nu'_\rs$ and $\tilde\nu$ are the frequencies measured
by the source and the absorber, respectively. Of course this red shift
is also the ratio of the particle energies measured
by the source and the absorber, respectively:}
z&=\frac{E'_\rs}{\tilde E}-1.
\end{align}
The red shift is actually a blue shift, when $z$ is negative.
\subsection{The rotating absorber}
Consider a light signal originating at the source at the time $t_\rs$.
According to the rest frame, this signal arrives at the detector
at the time $t_\ra$. The motion is inside the waveguide.
So the radial speed $\dot\rho$ satisfies
\begin{align}\label{3}
(\dot\rho)^2+\omega^2\,\rho^2&=c^2.
\intertext{Integrating this, one finds the difference between the arrival time and
the emission time:}
t_{\ra}-t_{\rs}&=\left|\int_{\zeta\,\rho_\rs}^{\rho_\ra}\frac{\rd\,\rho}{\sqrt{c^2-\omega^2\,\rho^2}}\right|.
\intertext{In the above, $\zeta$ is $1$ ($-1$) if the source and the absorber
are on the same (opposite) sides of the disc center. One arrives at,}
t_{\ra}-t_{\rs}&=T,
\intertext{where}
T&=\frac{|\sin^{-1}\beta_\ra-\zeta\,\sin^{-1}\beta_\rs|}{\omega}.\\
\beta&=\frac{\omega\,\rho}{c}.
\intertext{It is seen that $T$ itself is independent of time. So, if a tap is
emitted at the source from $t_{\rs\,1}$ to $t_{\rs\,2}$, and arrives at the absorber
from $t_{\ra\,1}$ to $t_{\ra\,2}$,}
t_{\ra\,2}-t_{\ra\,1}&=t_{\rs\,2}-t_{\rs\,1}.
\end{align}

According to the source, the signal is emitted from $t'_{\rs\,1}$
to $t'_{\rs\,2}$. The corresponding time difference (according to
the source) is the proper time corresponding to the world-line of the source
between these two events. The source is moving at a constant speed
$(\omega\,\rho_\rs)$. So,
\begin{align}
t'_{\rs\,2}-t'_{\rs\,1}&=\int_{t_{\rs\,1}}^{t_{\rs\,2}}\rd\,t\,\sqrt{1-\beta^2_\rs}.
\intertext{That is}
t'_{\rs\,2}-t'_{\rs\,1}&=\frac{t_{\rs\,2}-t_{\rs\,1}}{\gamma_\rs},
\intertext{where}
\gamma&=\frac{1}{\sqrt{1-\beta^2}}.
\intertext{Similarly, according to the absorber, the signal is arrived from $t'_{\ra\,1}$
to $t'_{\ra\,2}$. The corresponding time difference (according to
the absorber) is the proper time corresponding to the world-line of the absorber
between these two events. The absorber is moving at a constant speed $(\omega\,\rho_\ra)$. So,}
t'_{\ra\,2}-t'_{\ra\,1}&=\frac{t_{\ra\,2}-t_{\ra\,1}}{\gamma_\ra}.
\intertext{One then arrives at}
\frac{\Delta\,t'_\ra}{\Delta\,t'_\rs}&=\frac{\gamma_\rs}{\gamma_\ra}.
\intertext{The relation between $\nu'_\ra$ (the frequency that is received by the absorber)
and $\nu'_\rs$ (the frequency that is emitted by the source) is}
\frac{\nu'_\ra}{\nu'_\rs}&=\frac{\Delta\,t'_\rs}{\Delta\,t'_\ra}.
\intertext{So,}\label{15}
\frac{\nu'_\ra}{\nu'_\rs}&=\frac{\gamma_\ra}{\gamma_\rs}.
\end{align}
It is seen that the absorber receives a blue shifted (red shifted) frequency,
if its distance from the disc center is larger (smaller) that the distance
of the source from the disc center. It is to be noted that the time delay $T$
between the emission and absorption is irrelevant in this. Specifically,
this time delay is larger when the source and the absorber are on
the opposite sides of the disc center, compared to the case they are on
the same side (but with the same distances from the disc center).
But the frequency shift is the same.

Alternatively, one could calculate the energy of a photon according to the absorber.
A photon of the energy $E_\rs$ (according to the rest frame) is emitted at the source,
and is received at the absorber. When the photon arrives at the absorber,
its energy is $E_\ra$ (again according to the rest frame). The reason that $E_\ra$ is
different from $E_\rs$, is that the path of the photon (according to the rest frame)
is not a straight line. The momentum $P$ of the photon changes during its travel, and this
changes the energy of the photon as well: the photon is continuously bouncing
back from the wall of the waveguide. In the collision of a particle with
a wall moving with the velocity $(c\,\bm{\beta})$ the
change in the energy and momentum of
the particle are (as explained in the appendix) related through
\begin{align}\label{14}
\delta\,E&=c\,\bm{\beta}\cdot\delta\,\bm{P}.
\intertext{For the motion of the photon inside the waveguide, the energy and momentum
of the photon vary continuously: collisions produce infinitesimal
changes in the energy and momentum of the photon, and that the above becomes
a relation between the differentials of the energy and momentum:}
\rd\,E&=c\,\beta\,\hat{\bm{\phi}}\cdot\rd\,\bm{P},
\intertext{where $\hat{\bm{\rho}}$ and $\hat{\bm{\phi}}$ are the unit vectors in the radial
and angular directions, respectively. The above equation results in}
\rd\,E&=c\,\beta\,[\rd\,(\hat{\bm{\phi}}\cdot\bm{P})-\bm{P}\cdot\rd\,\hat{\bm{\phi}}],\nonumber\\
&=c\,\beta\,[\rd\,(\hat{\bm{\phi}}\cdot\bm{P})+(\hat{\bm{\rho}}\cdot\bm{P})\,\rd\,\phi].
\intertext{One has,}
c\,\hat{\bm{\phi}}\cdot\bm{P}&=\beta\,E.\\
c\,\hat{\bm{\rho}}\cdot\bm{P}&=\sqrt{1-\beta^2}\,E.
\intertext{Also, equation (\ref{3}) can be rewritten as}
\left(\frac{\rd\,\beta}{\rd\,\phi}\right)^2+\beta^2&=1,
\intertext{which results in}
\rd\,\phi&=\frac{\rd\,\beta}{\sqrt{1-\beta^2}}.
\intertext{One then arrives at}
\rd\,E&=\beta\,[\rd\,(\beta\,E)+E\,\rd\,\beta].
\intertext{That is,}
(1-\beta^2)\,\rd\,E&=E\,\rd\,(\beta^2).
\intertext{So,}
\frac{E_\ra}{E_\rs}&=\frac{1-\beta^2_\rs}{1-\beta^2_\ra}.
\intertext{Or,}\label{26}
\frac{E_\ra}{E_\rs}&=\frac{\gamma^2_\ra}{\gamma^2_\rs}.
\end{align}
The source and the absorber are moving with the velocities $(c\,\bm{\beta}_\rs)$ and
$(c\,\bm{\beta}_\ra)$, respectively. The Lorentz transformation relating the energy
of the photon according to the rest frame, to the energy of the photon according to
the source or the absorber, is
\begin{align}
E&=\gamma\,(E'+c\,\bm{\beta}\cdot\,\bm{P}').
\intertext{According to the source (absorber),
the photon is moving radially when it is at the location of the source (absorber).
So $\bm{P}'_\rs$ is normal to $\bm{\beta}_\rs$, and $\bm{P}'_\ra$ is normal to
$\bm{\beta}_\ra$. And one arrives at}
E_\rs&=\gamma_\rs\,E'_\rs.\\
E_\ra&=\gamma_\ra\,E'_\ra.
\intertext{So,}\label{30}
\frac{E'_\ra}{E'_\rs}&=\frac{\gamma_\ra}{\gamma_\rs}.
\intertext{This is the same shift (\ref{15}), which was found for the frequency.
For a rotating absorber,}
\tilde\nu&=\nu'_\ra.\\
\tilde E&=E'_\ra.\\
z&=\frac{\gamma_\rs}{\gamma_\ra}-1.
\end{align}
\subsection{The stationary absorber}
The quantities measured at the edge of the waveguide (a distance $r$
from the disc center), are denoted by the subscript $\re$.
The frequency $\nu_\rr$ (measured according to the rest frame)
is related to $\nu'_\re$, through the following Lorentz transformation.
\begin{align}
2\,\pi\,\nu_\rr&=\gamma_\re\,(2\,\pi\,\nu'_\re+c\,\bm{\beta}_\re\cdot\,\bm{k}'_\re).
\intertext{Where $\bm{k}'_\re$ is the wave vector received by the
point at the edge of the waveguide. This wave vector is radial,
hence normal to $\bm{\beta}_\re$. So,}
\nu_\rr&=\gamma_\re\,\nu'_\re.
\intertext{The frequency received by the absorber is $\nu_\rr$. And, using (\ref{15}),
one arrives at}\label{36}
\frac{\nu_\rr}{\nu'_\rs}&=\frac{\gamma^2_\re}{\gamma_\rs}.
\intertext{Of course}
\gamma_\re&=\frac{1}{\sqrt{1-\beta^2_\re}}.\\
\beta_\re&=\frac{\omega\,r}{c}.
\intertext{Again, one could arrive at this, using the energy of a photon emerging from
the waveguide. The energy of the photon at the detector is $E_\rr$, and
the following Lorentz transformation relates this to $E'_\re$:}
E_\rr&=\gamma_\re\,(E'_\re+c\,\bm{\beta}_\re\cdot\,\bm{P}'_\re).
\intertext{According to edge of the waveguide (which is rotating with the disc),
the photon is moving radially when it is at the edge.
So $\bm{P}'_\re$ is normal to $\bm{\beta}_\re$. Hence,}
E_\rr&=\gamma_\re\,E'_\re.
\intertext{And, using (\ref{30}), one arrives at}
\frac{E_\rr}{E'_\rs}&=\frac{\gamma^2_\re}{\gamma_\rs}.
\intertext{It is seen that this is consistent with (\ref{36}). For a stationary absorber,}
\tilde\nu&=\nu_\rr.\\
\tilde E&=E_\rr.\\
z&=\frac{\gamma_\rs}{\gamma^2_\re}-1.
\end{align}
\section{The detected particles}
In the absence of the absorber, it is expected that $N_\rs$ particles be detected
at the detector. The emitted particles have an energy spectrum $\Upsilon$: The number of emitted particles
with energy around $E$ and in an interval of $(\Delta\,E)$,
is $[(N_\rs\,\Delta E)\,\Upsilon(E)]$. Obviously,
\begin{align}
1&=\int\rd\,E\,\Upsilon(E).
\intertext{The probability that a particle with the energy $E$ is absorbed, is
$[A(E)]$. For a particle emitted with the energy $E'_\rs$ (as measured by the source),
the energy that is measured by the absorber is $\tilde E$. The number of particle
which are detected is denoted by $N_\rd$. Assuming no loss, apart from the absorption, one has}
\frac{N_\rd}{N_\rs}&=1-\int\rd\,E'_\rs\,\Upsilon(E'_\rs)\,A\left(\frac{E'_\rs}{1+z}\right).
\intertext{Or,}\label{47}
\frac{N_\rd}{N_\rs}&=1-(1+z)\,\int\rd\,\tilde E\,\Upsilon[(1+z)\,\tilde E]\,A(\tilde E).
\intertext{A special case is when the absorption function $A$ is sharply peaked
around some $E_\rm$. In that case, (\ref{47}) can be approximated by}
\frac{N_\rd}{N_\rs}&=1-(1+z)\,\Upsilon[(1+z)\,E_\rm]\,\int\rd\tilde E\,A(\tilde E).
\intertext{The integral on the right-hand side is independent of
the state of the absorber. Denoting that by $B$, one arrives at}
\frac{N_\rd}{N_\rs}&=1-(1+z)\,B\,\Upsilon[(1+z)\,E_\rm].
\intertext{In almost are practical situations, $|z|$ is much smaller than $1$. So,}
\frac{N_\rd}{N_\rs}&=1-B\,\Upsilon(E_\rm)-z\,B\,[\Upsilon(E_\rm)+E_\rm\,(\rD\,\Upsilon)(E_\rm)]+\cdots,
\intertext{where $(\rD\,\mathfrak{X})$ is the derivative of $\mathfrak{X}$. Also,}
z&=\frac{\sigma\,\omega^2}{c^2}+O(c^{-4}),
\intertext{where $\sigma$ is a constant of the dimension length squared. The final result is then}
\frac{N_\rd}{N_\rs}&=1-B\,\Upsilon(E_\rm)
-\{B\,[\Upsilon(E_\rm)+E_\rm\,(\rD\,\Upsilon)(E_\rm)]\}\,\frac{\sigma\,\omega^2}{c^2}+O(c^{-4}).
\end{align}
Some special cases are discussed below.
\subsection{The rotating absorber}
The absorber is attached to the disc, and rotates with the disc:
\begin{align}
\sigma&=\frac{\rho^2_\rs-\rho^2_\ra}{2}.\\
\intertext{In particular, if the disc center is half way between the source and the absorber:}
\rho_\rs&=\rho_\ra,
\intertext{then}
\sigma&=0.
\intertext{And if the source is on the disc center:}
\rho_\rs&=0,
\intertext{then}
\sigma&=-\frac{\rho^2_\ra}{2}.
\end{align}
\subsection{The stationary absorber}
The absorber is just outside the disc, stationary according to the rest frame:
\begin{align}
\sigma&=\frac{\rho^2_\rs-2\,\rho^2_\re}{2}.
\intertext{In particluar, if the source is on the disc center:}
\rho_\rs&=0,
\intertext{then,}
\sigma&=-\rho^2_\re.
\end{align}
\section{Concluding remarks}
The relativistic frequency shift caused by rotating waveguide was studied.
In the setup, a radiation source is in a waveguide, which is on a diagonal
of a rotating disc. Both the waveguide and the source rotate with the disc.
An absorber absorbs some of the photons, decreasing the number of the photons
which reach a detector. The absorption fraction depends on the frequency of
the photons according to the absorber, and this can be used as a method to
measure possible frequency shifts from the source to the detector.
Among other things, it was shown even if both the source and the absorber
are stationary, there is still a nonzero shift from the source to the absorber.
The reason is that according to the rest frame, in which the disc is rotating,
the photon is moving on a curved path, hence its energy (frequency)
is constantly changing.

Based on the results, various M\"{o}ssbauer rotor experiments
can be designated as further tests of special relativity.
\section{Appendix: Relativistic collision with a wall}
Consider a two body collision. The conservation relations are
\begin{align}
\mathcal{E}+E&=\mathcal{E}_\rf+E_\rf.\\
\bm{\mathcal{P}}+\bm{P}&=\bm{\mathcal{P}}_\rf+\bm{P}_\rf.
\intertext{Quantities corresponding to the second (later large) body are denoted by
calligraphic letters, and the subscript $\rf$ refers to
the quantities after the collision. The above equations result in}
\mathcal{E}^2_\rf-c^2\,\bm{\mathcal{P}}_\rf\cdot\bm{\mathcal{P}}_\rf&=
(\mathcal{E}+E-E_\rf)^2-c^2\,(\bm{\mathcal{P}}+\bm{P}-\bm{P}_\rf)\cdot(\bm{\mathcal{P}}+\bm{P}-\bm{P}_\rf).
\intertext{That is,}
\mathcal{E}^2_\rf-c^2\,\bm{\mathcal{P}}_\rf\cdot\bm{\mathcal{P}}_\rf&=
\mathcal{E}^2-c^2\,\bm{\mathcal{P}}\cdot\bm{\mathcal{P}}\nonumber\\
&\quad\,+2\,[\mathcal{E}\,(E-E_\rf)-c^2\,\bm{\mathcal{P}}\cdot\,(\bm{P}-\bm{P}_\rf)]\nonumber\\
&\quad\,+(E-E_\rf)^2-c^2\,(\bm{P}-\bm{P}_\rf)\cdot(\bm{P}-\bm{P}_\rf).
\end{align}
Using the mass shell relation for the second body, one arrives at
\begin{align}
2\,(\mathcal{E}\,\delta\,E-c^2\,\bm{\mathcal{P}}\cdot\,\delta\,\bm{P})&=
(\delta\,E)^2-c^2\,(\delta\,\bm{P})\cdot(\delta\,\bm{P}),
\intertext{where}
\delta\,\mathfrak{X}&=\mathfrak{X}_\rf-\mathfrak{X}.
\intertext{One has}
c\,\bm{\mathcal{P}}&=\bm{\beta}\,\mathcal{E}.
\intertext{Where $(c\,\bm{\beta})$ is the velocity of the second body (before the collision). So,}
\delta\,E-c\,\bm{\beta}\cdot\,\delta\,\bm{P}&=
\frac{(\delta\,E)^2-c^2\,(\delta\,\bm{P})\cdot(\delta\,\bm{P})}{2\,\mathcal{E}}.
\intertext{The collision with a {\em wall} corresponds to an infinite $\mathcal{E}$.
In that case,}
\delta\,E-c\,\bm{\beta}\cdot\,\delta\,\bm{P}&=0.
\end{align}
This is the same as (\ref{14}).
\\[\baselineskip]
\noindent\textbf{Acknowledgement}: This work was
supported by the Research Council of the Alzahra University.

\end{document}